\documentclass[aps,prb,twocolumn,showpacs,superscriptaddress, preprintnumbers]{revtex4-1}
\usepackage{latexsym}
\usepackage{amssymb}
\usepackage{graphicx}
\usepackage{amsmath}
\usepackage{bm}
\usepackage[colorlinks,
          linkcolor=black,
            citecolor=black,
            urlcolor=blue
           ]{hyperref}
\usepackage{verbatim}
\usepackage{mathrsfs}
\usepackage{extarrows}
\usepackage{comment}
\usepackage{mathtools,slashed}
\usepackage{soul}
\usepackage[toc,page]{appendix}
\usepackage[vcentermath]{youngtab}
\usepackage{multirow}
\usepackage{atbegshi,picture}
\usepackage{lipsum}
\usepackage{bbm}

\usepackage{ulem}

\begin{document}

\title{Geometric approach to Lieb-Schultz-Mattis theorem without translation symmetry under inversion or rotation symmetry}

\author{Yuan Yao}
\email{yuan.yao@riken.jp}
\affiliation{Condensed Matter Theory Laboratory, RIKEN CPR, Wako, Saitama 351-0198, Japan}

\author{Akira Furusaki}
\affiliation{Condensed Matter Theory Laboratory, RIKEN CPR, Wako, Saitama 351-0198, Japan}
\affiliation{Quantum Matter Theory Research Team, RIKEN CEMS, Wako, Saitama 351-0198, Japan}

\date{\today}

\begin{abstract}

We propose a geometric {approach to Lieb-Schultz-Mattis theorem for} quantum many-body systems with discrete spin-rotation symmetries and lattice inversion or rotation symmetry, but without translation symmetry assumed.
Under symmetry-twisting on a $(d-1)$-dimensional plane,
we find that any $d$-dimensional inversion-symmetric spin system possesses a doubly degenerate spectrum when it hosts a half-integer spin at the inversion-symmetric point.
We also show that any rotation-symmetric generalized spin model with a projective representation at the rotation center has a similar degeneracy under symmetry-twisting.
We argue that these degeneracies imply that {a unique symmetric gapped ground state that is smoothly connected to product states} is forbidden in the original untwisted systems
---  generalized inversional/rotational Lieb-Schultz-Mattis theorems without lattice translation symmetry imposed.
The traditional Lieb-Schultz-Mattis theorems with translations also fit in the proposed framework.

\end{abstract}

\maketitle

\section{Introduction}
Strongly interacting many-body systems are central topics in condensed-matter and statistical physics.
An important concept in the study of these systems is the Lieb-Schultz-Mattis (LSM) theorem~\cite{Lieb:1961aa} and its generalizations~\cite{Affleck:1986aa,OYA1997,Oshikawa:2000aa, Hastings:2004ab} that 
state an \textit{ingappability} --- inability of having a unique gapped ground state that {belongs to the phase of a trivial product state} --- of the systems respecting U$(1)$ and translation symmetries with a fractional filling.
Recent works study the ingappabilities from the interplay between translations and other symmetries, e.g., SU$(2)$~\cite{Furuya:2017aa,Metlitski:2018aa} and SU$(N)$~\cite{Affleck:1986aa,Yao:2019aa} symmetries, or even discrete subgroup symmetries~\cite{Chen-Gu-Wen_classification2010,Fuji-SymmetryProtection-PRB2016,Watanabe:2015aa,Ogata:2018aa,Ogata:2020aa,Yao:2021aa}.

The lattice translation symmetry is essential in these LSM-type theorems.
Several generalizations are recently proposed to systems with other lattice symmetries than translations~\cite{Watanabe:2015aa,Po:2017aa,Huang:2017aa,Else:2020aa,Ogata:2018aa,Ogata:2020aa}.
Such extensions have been made in spin systems with rotation or inversion symmetry in low dimensions by employing lattice homotopy principles~\cite{Po:2017aa,Else:2020aa} {and} even proven in a rigorous manner in one dimension~\cite{Ogata:2018aa,Ogata:2020aa}.
{The} ingappabilities of these systems are also related to quantum anomaly of field theories~\cite{Cheng:2016aa,Metlitski:2018aa,Else:2020aa}.
Nevertheless,
higher dimensional generalizations are not well established and require further studies, e.g., inversions beyond one and two dimensions in a unifying way.
Moreover,
convincing lattice-based arguments are still lacking for systems with general lattice rotational and inversional symmetries.

In this work,
we propose a geometric picture to study LSM-type (in)gappabilities for generic lattice systems without translation symmetries.
As concrete {examples},
we first focus on spin systems with discrete spin-rotation symmetries (rather than the full SO$(3)$ for generality) and site-centered inversions in arbitrary dimensions,
and general lattice rotations in two dimensions.
We consider closed geometry of finite lattice systems by identifying boundary spins, consistently with the lattice symmetries. 
We then twist the boundary condition using spin-rotation symmetries in a certain geometric pattern.
Assuming that bulk properties are insensitive to such twisting at the boundary~\cite{Watanabe:2018aa,Yao:2020PRX,Yao:2021aa}, we can extract ingappabilities of the original (untwisted) system from the interplay between the geometric pattern and lattice symmetries.
It is expected to apply to broader classes of systems, e.g., with nonsymmorphic symmetries~\cite{Parameswaran:2013,Watanabe:2015aa}.

\section{Geometric patterns of symmetry-twisted boundary conditions}
\begin{figure}[t]
\centering
\includegraphics[width=8cm,pagebox=cropbox,clip]{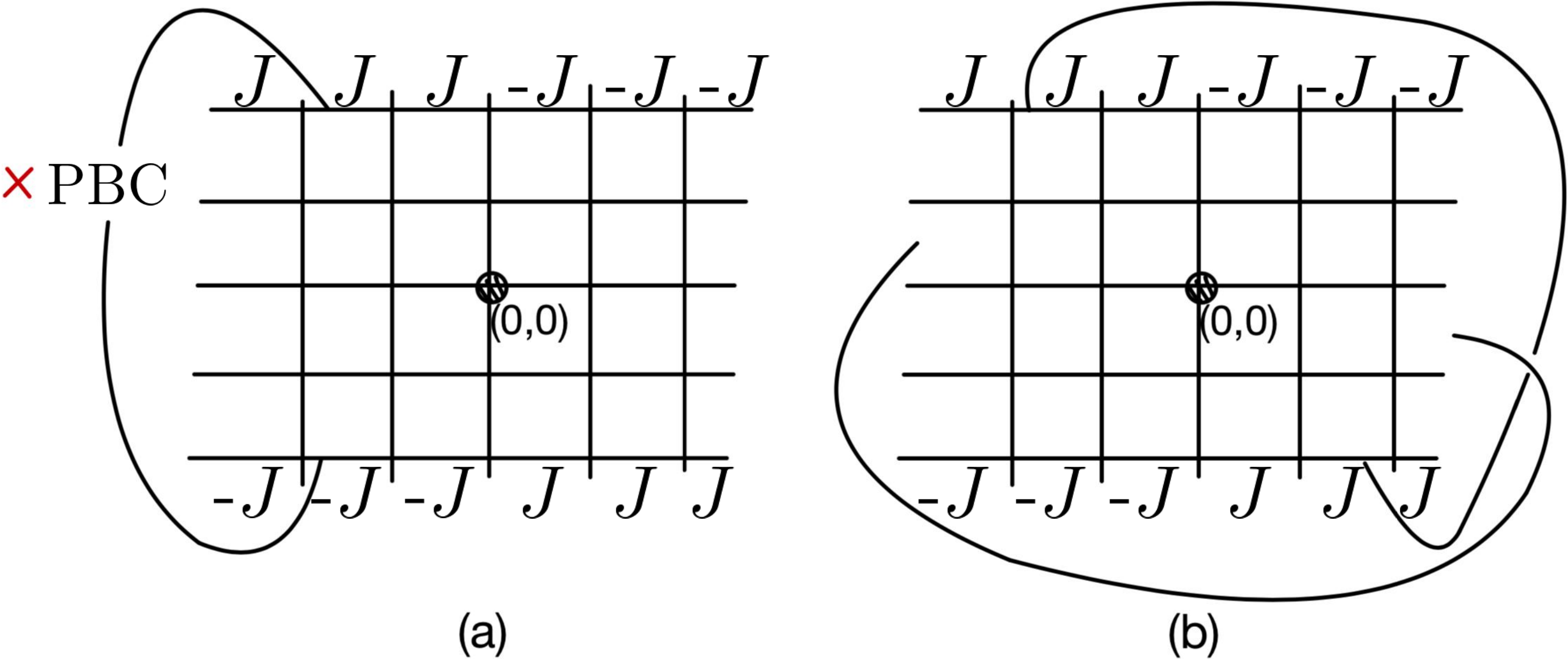}
\caption{(a)~An inversion-symmetric Hamiltonian with coupling constants $\pm J$ has a boundary that cannot be closed by PBC. (b)~The boundary closing by identifications $\vec{r}\sim-\vec{r}$ is compatible with any inversion-symmetric Hamiltonian.
The resultant manifold in the continuum limit is a real projective plane $\mathbb{RP}^2$.
}
\label{PBC_RBC}
\end{figure}

{{An important indicator of the LSM-type ingappabilities is} the ground-state degeneracies of {a many-body} system on a closed lattice {that take values larger than unity in the presence of} SSB or fractionalization.}
Quite often, 
periodic boundary conditions (PBCs) are {assumed for} Hamiltonians on square lattices with sizes $L_i$ by identifying sites $\vec{r}_i\sim\vec{r}_i+L_i$.
However,
in the following discussion, 
except for certain symmetry requirements, 
we do not specify any concrete form of Hamiltonian in advance,
{and we will choose boundary conditions that are} compatible with all possible Hamiltonians respecting the required symmetries.
For instance,
when the imposed symmetry is the inversion $\vec{r}\rightarrow-\vec{r}$ about the origin $\vec{r}=\vec{0}$,
an inversion-symmetric Hamiltonian that cannot be closed by PBC is shown in FIG.~\ref{PBC_RBC}~(a); PBC is inapplicable when only the inversion is imposed~\footnote{If we also impose lattice translation symmetries, then PBC is permitted.  It is consistent since the Hamiltonian in FIG.~\ref{PBC_RBC} breaks translations thereby out of our consideration.}.
{Instead, the boundary closing shown in FIG.~\ref{PBC_RBC}~(b) is compatible with the inversion symmetry.}

\begin{figure}[t]
\centering
\includegraphics[width=8cm,pagebox=cropbox,clip]{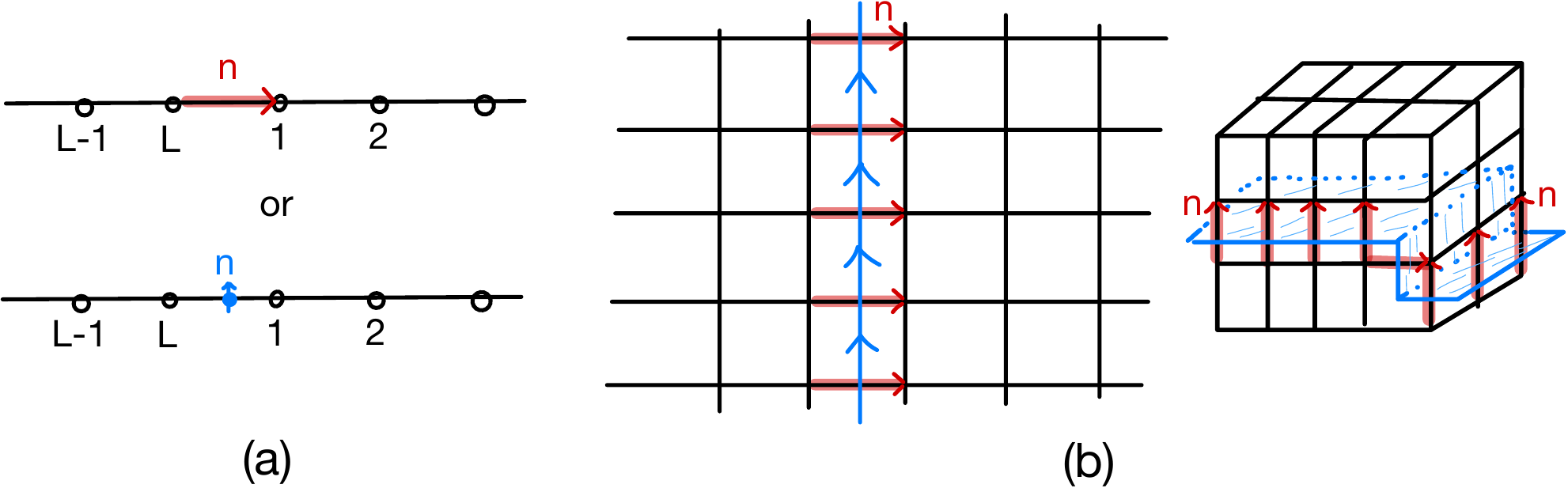}
\caption{Two equivalent representations of $n\in\mathbb{Z}_m$ twisting in {one dimension~(a) and higher dimensions~(b).}
We will use either of them whichever is convenient.
}
\label{twisting_dual}
\end{figure}

When the system possesses internal symmetries, such as U$(1)$ or discrete $\mathbb{Z}_m$ symmetry,
we can ``mix'' them into the boundary condition as follows.
Let us take $\mathbb{Z}_m$ twisting in tight-binding models with charged operator $c_i^\dagger$ for illustration.
In one dimension,
{angle-$(2\pi n/m)$} twisting with an integer $n$ can be {introduced} by substituting the boundary hopping term $\left[c_Lc_1^\dagger+\text{h.c.}\right]$ with $\left[\exp(i2\pi n/m)c_Lc_1^\dagger+\text{h.c.}\right]$.
For general coupling terms involving multiple sites crossing the bond between sites $L$ and $1$,
we can apply the U$(1)$ transformation $c_j^\dagger\rightarrow\exp(i2\pi n/m)c_j^\dagger$ \textit{only} on the sites $j$ on the ``right'' of the bond~\footnote{Because the range $l$ of interaction terms $l\ll L$, there is a well-defined ``left/right'' only locally around the boundary.}.
This twisting has two equivalent geometric presentations in FIG.~\ref{twisting_dual}~(a).
The first one is the obvious type, where a twisted boundary bond is marked in bold with an arrow pointing to the ``right''.
The second type is the dual one: an arrowed dot is drawn on this bond to indicate that interaction terms across this dot get twisted by the phase specified with the integer $n$.
These two dual pictures are almost the same in one dimension,
but their differences will be clear in higher dimensions, making one of them more convenient than the other depending on situations.

In two dimensions,
we can similarly twist all bonds crossing the boundary line by the internal symmetries as shown in FIG.~\ref{twisting_dual}~(b).
The dual picture is exactly the boundary line which transversally intersects those bond centers, and the line is labelled by $n\in\mathbb{Z}_m$ with oriented arrows.
In general $d$ dimensions,
we twist the bonds intersected by the $(d-1)$-dimensional boundary hypersurface.
The dual picture has a certain hypersurface with an orientation and its symmetry label, which is a special case of the Poincar\'e duality.

\begin{figure}[t]
\centering
\includegraphics[width=8.5cm,pagebox=cropbox,clip]{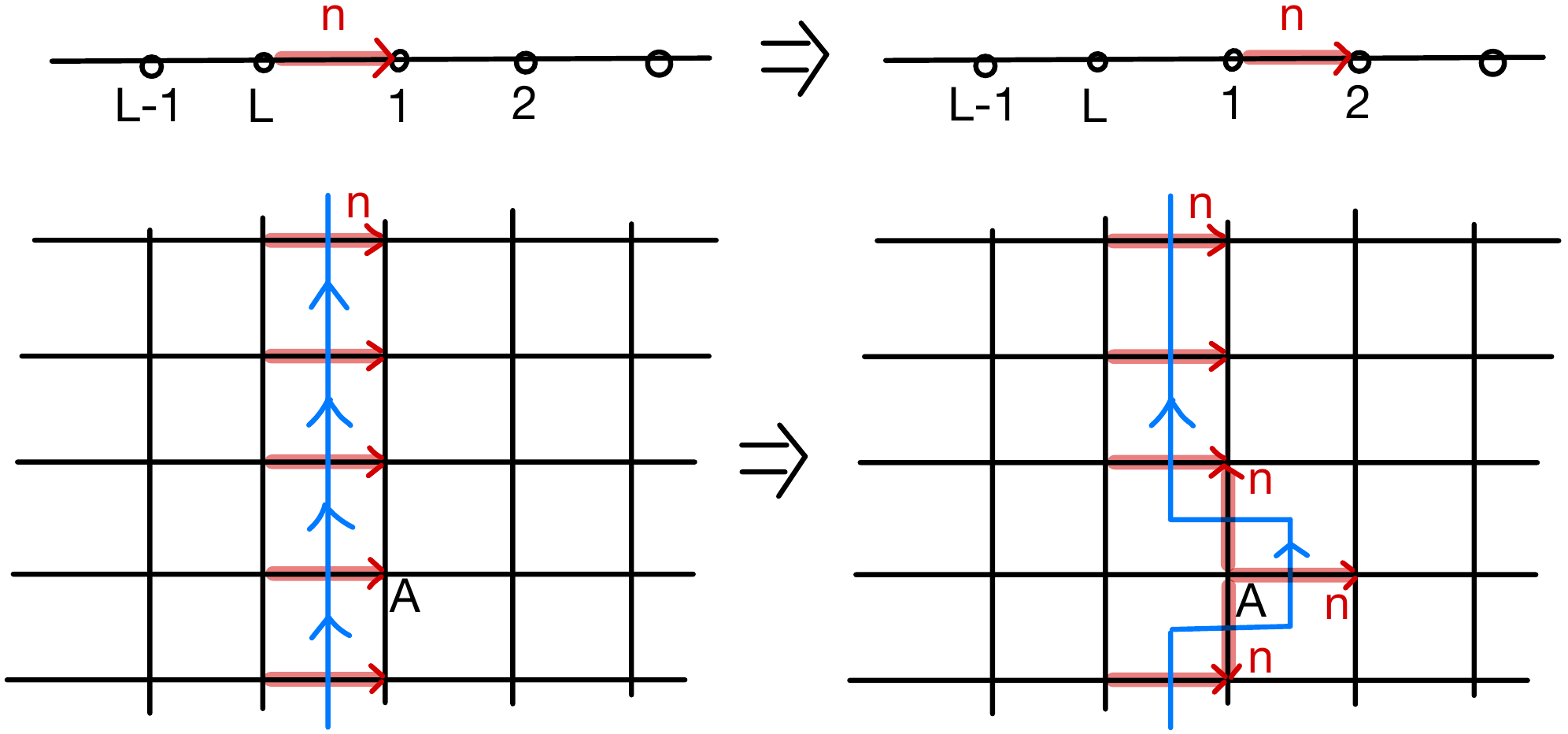}
\caption{The $n\in\mathbb{Z}_m$ twisting can be moved/deformed by a gauge transformation: $\mathcal{H}_\text{tw}\rightarrow U^{-1}\mathcal{H}_\text{tw}U$ with $U=\exp(i2\pi c^\dagger_{1\text{(A)}}c_{1\text{(A)}}n/m)$.
}
\label{flatness}
\end{figure}

Since the boundary of the boundary is always empty,
the boundary line/face (transversal to all the twisted bonds) must be closed themselves, i.e., closed loops/surfaces.
It exactly means that the total net symmetry twisting around any plaquette must be zero.
In the dual picture of symmetry twisting,
we can {argue} that this mixture of twisting into the boundary condition does not introduce any boundary modes at the twisted bonds.
It is a direct consequence of the following gauge invariance.
We can use a gauge freedom to rename the charged operator by a unitary operator.
For example in one dimension, the transformation
$c^\dagger_1\rightarrow c^\dagger_1\exp(-i2\pi n/m)$ with a unitary operator $U=\exp(i2\pi c^\dagger_1c_1n/m )$ undoes the earlier twisting of the bond between sites $L$ and $1$.
However, it creates a new twisting between sites $1$ and $2$,
so effectively the twisting is moved by one site as shown in the upper figure of FIG.~\ref{flatness}.
Thus we can move the twisting freely by unitary transformations,
which implies the absence of physical boundary modes there.

The above argument using gauge degrees of freedom applies to higher dimensions as shown in FIG.~\ref{flatness};
the loop/surface transversal to the twisted bonds can be deformed arbitrarily by local unitary transformation, thereby unchanging any energy-spectrum property.
It should be noted that
such deformations are performed only locally in that they cannot make a non-contractible loop into a contractible one.
These gauge properties of twisting induce the concept of bulk insensitivity, i.e.,
the insensitibity of the LSM-type gappability to the symmetry twisting. 
In other words,
if the system has a unique symmetric gapped ground state {that is in the same phase as trivial product states},
%without SSB {or} fractionalization before symmetry twisting,
it will still have a unique symmetric gapped ground state after the symmetry twisting.
The bulk insensitivity is physically reasonable although not proven in general.
In fact, 
it is only proven under a certain assumption on the first excited state in one dimension~\cite{Watanabe:2018aa} and justified by a quantum-transfer-matrix formalism~\cite{Yao:2021aa} in higher dimensions.
In the following discussion,
we will assume the insensitivity to twisted boundary conditions to derive LSM-type theorems.
% {(without considering invertible or intrinsic topological orders).}

{
A comment is here in order.
When we impose the boundary condition of Fig.~\ref{PBC_RBC}~(b) on a ground state in an invertible or intrinsic topological ordered phase~\cite{WenRMP2017} (e.g., quantum Hall phases or its higher-dimensional generalizations) that has (chiral) boundary modes,
%which still has a potential inconsistency when the bulk is in an invertible topological ordered phase,
%which is characterized by chiral boundary modes,
%e.g., quantum Hall phases or its higher dimensional generalizations.
the chiral boundary modes remain gapless because two modes with the same chirality cannot be gapped.
Therefore, our theory with inversion symmetry does not apply to invertible/intrinsic topological ordered phases.
%we will always assume that the invertible or intrinsic topological order is absent so that such a boundary closing can be done.
}

\section{Inversional LSM theorems}
\begin{figure}[t]
\centering
\includegraphics[width=8.5cm,pagebox=cropbox,clip]{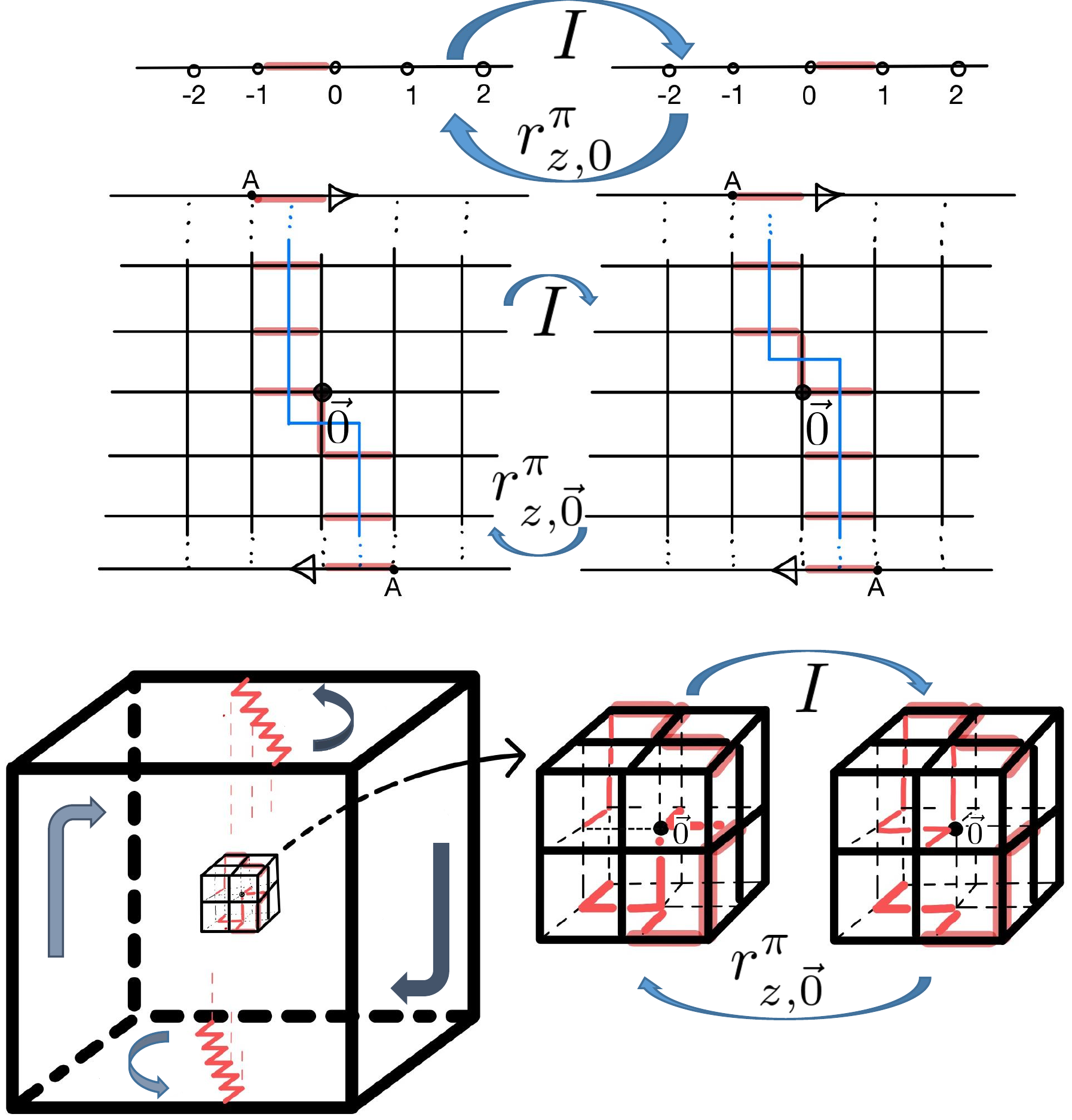}
\caption{The twisted Hamiltonians $\mathcal{H}_\text{tw}$ on the left side are transformed by inversions to $I^{-1}\mathcal{H}_\text{tw}I$ on the right side.
$I^{-1}\mathcal{H}_\text{tw}I$ can be transformed back by the gauge transformation: $(r^\pi_{z,0})^{-1}(I^{-1}\mathcal{H}_\text{tw}I)r^\pi_{z,0}=\mathcal{H}_\text{tw}$, which implies $\mathcal{H}_\text{tw}$ is symmetric under $\tilde{I}$ in Eq.~(\ref{inv}).
The sites on the boundary are identified by $\vec{r}\sim-\vec{r}$,
so the continuum limit of lattice in $d$ dimensions is a real projective hyperplane $\mathbb{RP}^d$.
}
\label{inversion_LSM}
\end{figure}

Let us consider a spin chain $\mathcal{H}_0$ with length $L$ respecting {a site-centered inversion symmetry
\begin{eqnarray}
I^{-1}\vec{S}_jI=\vec{S}_{-j},
\end{eqnarray}
and} $\mathbb{Z}_2\times\mathbb{Z}_2$ discrete spin-rotation symmetry: 
\begin{eqnarray}\label{spin-rotation}
&&R^\pi_x=\exp\left(i\pi \sum^L_{j=1}S^x_j\right);\,\,R^\pi_z=\exp\left(i\pi \sum^L_{j=1}S^z_j\right),
\end{eqnarray}
where the spin operator $\vec{S}_j\equiv[S^x_j,S^y_j,S^z_j]$ is not necessarily of the same spin representation for all site $j$, except for those related by the inversion symmetry, and the spin of $\vec{S}_{j=0}$ at the inversion center is $s$.
Considering the inversion symmetry,
we can close the chain by identifying the boundary sites at the two ends that are related by inversion to each other.
%\blue{This boundary closing results in a chain of an even number of sites,
%free of Kramers degeneracy well-known in zero-dimensional quantum mechanics.}
%If, otherwise, there are an odd number of sites, the spectrum must be degenerate once the spin at the inversion center has half-integer spins by a Kramers-type argument in quantum mechanics. However, such a degeneracy due to a single-body effect is already well-known and not in our interest, so we need to avoid this special situation.}.
We then twist the chain by $\mathbb{Z}_2$ symmetry: $R^\pi_z$.
Since the twisting position can be changed freely by gauge invariance,
we put the twisting on the bond between the sites $-1$ and $0$ and denote the twisted Hamiltonian as $\mathcal{H}_\text{tw}$.
A typical example is the $XYZ$ model:
\begin{eqnarray}
\mathcal{H}_\text{tw}^{XYZ}&=&\sum_{j\neq0}h_j^{XYZ}-J_xS^x_{-1}S^x_0\!-\!J_yS^y_{-1}S^y_0+J_zS^z_{-1}S^z_0,\nonumber
\end{eqnarray}
where $h^{XYZ}_j\equiv \sum_{\alpha=x,y,z} J_\alpha S^\alpha_{j-1}S^\alpha_{j}$.
The twisted Hamiltonian $\mathcal{H}_\text{tw}$ is represented by FIG.~\ref{inversion_LSM}~(up-left), where the arrows are irrelevant in case of $\mathbb{Z}_2$-twisting.
It still respects $\mathbb{Z}_2\times\mathbb{Z}_2$ symmetry but explicitly breaks the inversion symmetry $I$ since the twisting is moved to the bond between sites $0$ and $1$ by $I$ in FIG.~\ref{inversion_LSM}~(up-right).
Nevertheless,
we can map it back by a gauge transformation $r^\pi_{z,0}\equiv\exp(i\pi S^z_{j=0})$.
It implies that $\mathcal{H}_\text{tw}$ respects the following modified inversion symmetry:
\begin{eqnarray}\label{inv}
\tilde{I}\equiv I\cdot r^\pi_{z,0};\,\,[\tilde{I},\mathcal{H}_\text{tw}]=0;\,\,\tilde{I}R^\pi_x=(-1)^{2s}R^\pi_x\tilde{I},
\end{eqnarray}
and $\mathcal{H}_\text{tw}$ still respects $R^\pi_{x,z}$.
%\begin{eqnarray}\label{commu}
%\tilde{I}R^\pi_x=(-1)^{2s}R^\pi_x\tilde{I},
%\end{eqnarray}
Here the factor $(-1)^{2s}$ comes from the commutator
\begin{eqnarray}
r^\pi_{z,0}r^\pi_{x,0}=(-1)^{2s}r^\pi_{x,0}r^\pi_{z,0}.
\end{eqnarray}
It implies that $\mathcal{H}_\text{tw}$ must have a doubly degenerate energy spectrum if $s=1/2,3/2,\cdots$, i.e., a half-integer spin at the origin.
Thus, the original Hamiltonian $\mathcal{H}_0$ cannot have a unique gapped ground state when $s=1/2,3/2,\cdots$ --- either a gapless or degenerate ground states should result in the thermodynamic limit, because, otherwise, the bulk insensitivity would mean that $\mathcal{H}_\text{tw}$ could also have a unique gapped ground state that contradicts the above double degeneracy.
{Note that the two-fold degeneracy discussed above is not due to the Kramers theorem (applicable only when the total spin being a half integer) but comes from the projective representation of $\vec{S}_0$.}

This approach {can be generalized to} arbitrary higher dimensions with $\mathbb{Z}_2\times\mathbb{Z}_2$ and inversion $I$ about the origin $\vec{r}\rightarrow-\vec{r}$,
but the boundary closing needs more consideration.
For instance, in two dimensions, PBC is not consistent as shown in FIG.~\ref{PBC_RBC}~(a), and the only sensible {way of closing is identifying} boundary sites $\vec{S}_{\vec{r}}\sim\vec{S}_{-\vec{r}}$ in FIG.~\ref{PBC_RBC}~(b) that is compatible with any inversion-symmetric Hamiltonian $\mathcal{H}_0$.
Then we twist $\mathcal{H}_0$ by $\mathbb{Z}_2: R^z_\pi$ as in FIG.~\ref{inversion_LSM}~(middle-left) with a series of twisted bonds or, in the dual picture, a closed loop.
The twisted Hamiltonian $\mathcal{H}_\text{tw}$ softly breaks $I$ and has the modified inversion $\tilde{I}$ in the same form as Eq.~(\ref{inv}).
Therefore, by {employing a} similar argument as above, {we see that}
$\mathcal{H}_0$ cannot have a unique gapped ground state when the spin $s$ at $\vec{r}=0$ is half integer.

The twisted Hamiltonian $\mathcal{H}_\text{tw}$ in three dimensions is sketched in FIG.~\ref{inversion_LSM}~(bottom). 
The twisting satisfies the {closed-form} condition that the total twisting around any plaquette is zero.
The same result {as in the lower-dimensional cases} follows since $\mathcal{H}_\text{tw}$ respects $\tilde{I}$.
The geometric situation in arbitrary dimensions is that when we {perform the inversion transformation, the spin at $\vec{r}=0$ is wrapped by the hypersurface (in the dual picture) spanned by the centers of twisted bonds together with the hypersurface inverted by $I$}.
{The} wrapping is exactly cancelled by the gauge transformation $r^\pi_{z,0}$, which gives the commutator as in Eq.~(\ref{inv}).
{It follows that $\mathcal{H}_0$ respecting $\mathbb{Z}_2\times\mathbb{Z}_2$ and $I$ is LSM-type ingappable in arbitrary dimensions when $s$ is not an integer.}

\section{Rotational LSM theorems}
In two dimensions,
the inversion $\vec{r}\rightarrow-\vec{r}$ is equivalent to lattice rotation by $180$ degrees.
It is natural to generalize $I$ to be $N$-fold rotations $C_N$ generated by a lattice rotation by $2\pi/N$ around the origin $\vec{r}=\vec{0}$, e.g., $C_3$ on a {honeycomb} lattice:
\begin{eqnarray}
C_N: \vec{r}\rightarrow\left[\begin{array}{cc}\cos(2\pi/N)&-\sin(2\pi/N)\\
\sin(2\pi/N)&\cos(2\pi/N)\end{array}\right]\vec{r}.
\end{eqnarray}
However,
even {with} translation symmetries,
spin-$1/2$ systems on the honeycomb lattice admits a unique gapped ground state with $C_3$ and SO$(3)$ spin-rotation symmetries~\cite{Kimchi-featureless-honeycomb,Metlitski:2018aa}.
It motivates us to consider SU$(3)$ ``spin'' degrees of freedom or generally SU$(N)$ ``spins'' obeying {an} su$(N)$ algebra~\cite{Affleck:1987aa,Affleck:1988aa}.
However,
the only {condition to be used here} is that there is a discrete ``spin''-rotation symmetry $\mathbb{Z}_N\times\mathbb{Z}_N$ (known as shift symmetries~\cite{Shimizu:2018aa,Tanizaki:2017aa}) by 
\begin{eqnarray}
V_N=\prod_{\vec{r}}v_{N,\vec{r}}\text{ and }W_N=\prod_{\vec{r}}w_{N,\vec{r}},
\end{eqnarray}
which generalize $R^\pi_z$ and $R^\pi_x$ {of} the $N=2$ case in Eq.~(\ref{spin-rotation})~\footnote{The only basis-independent property of $v_{N,\vec{r}}$ and $w_{N,\vec{r}}$ is the commutator such as Eq.~(\ref{comm_suN}). A matrix expression of $v_{N,\vec{r}}$ and $w_{N,\vec{r}}$ can be found, for example, in Eq.~(2.2) in Ref.~\cite{Yao:2021ab}.}.
There is a number $b$ called Young-tableau box number~\cite{Georgi:2000}, {analogous to the spin $s$ above},
to characterize the SU$(N)$ ``spin'' at the origin by the following commutator:
\begin{eqnarray}\label{comm_suN}
v_{N,\vec{r}=0}w_{N,\vec{r}=0}=\exp\!\left(i\frac{2\pi b}{N}\right)w_{N,\vec{r}=0}v_{N,\vec{r}=0}.
\end{eqnarray}
Thus $b=2s$ when $N=2$ and we do not need detailed forms of $v_{N,\vec{r}}$ and $w_{N,\vec{r}}$.

\begin{figure}[t]
\centering
\includegraphics[width=8.8cm,pagebox=cropbox,clip]{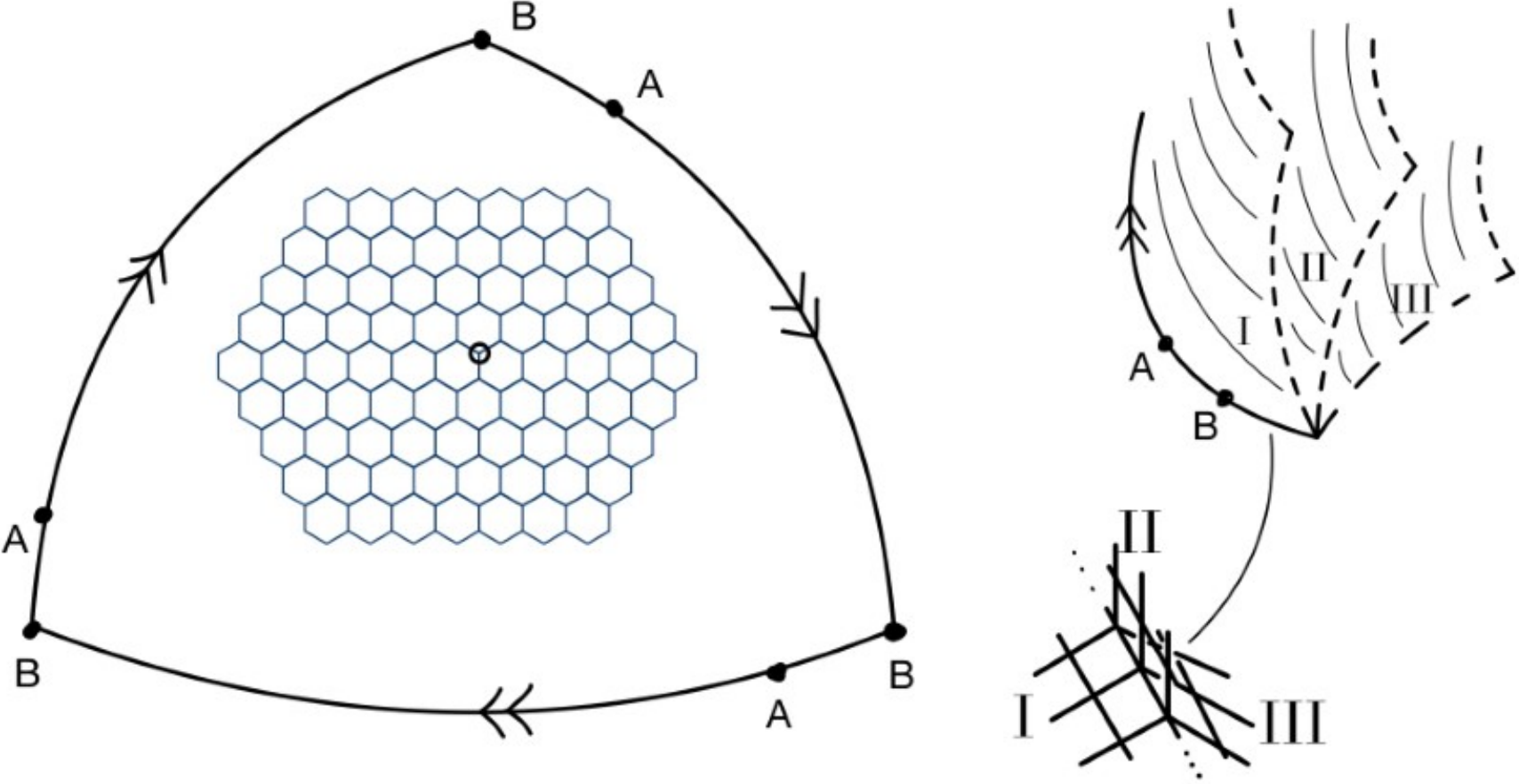}
\caption{The only boundary closing compatible with any $C_3$-symmetric Hamiltonian is to identify three boundary edges as above.
The continuum limit is a pseudo-projective plane of order $3$ with a three-page intersection~(up-right) which is realized by a simple identification of three edges at the lattice scale~(bottom-right).
}
\label{PPP_3}
\end{figure}

\begin{figure}[t]
\centering
\includegraphics[width=8.8cm,pagebox=cropbox,clip]{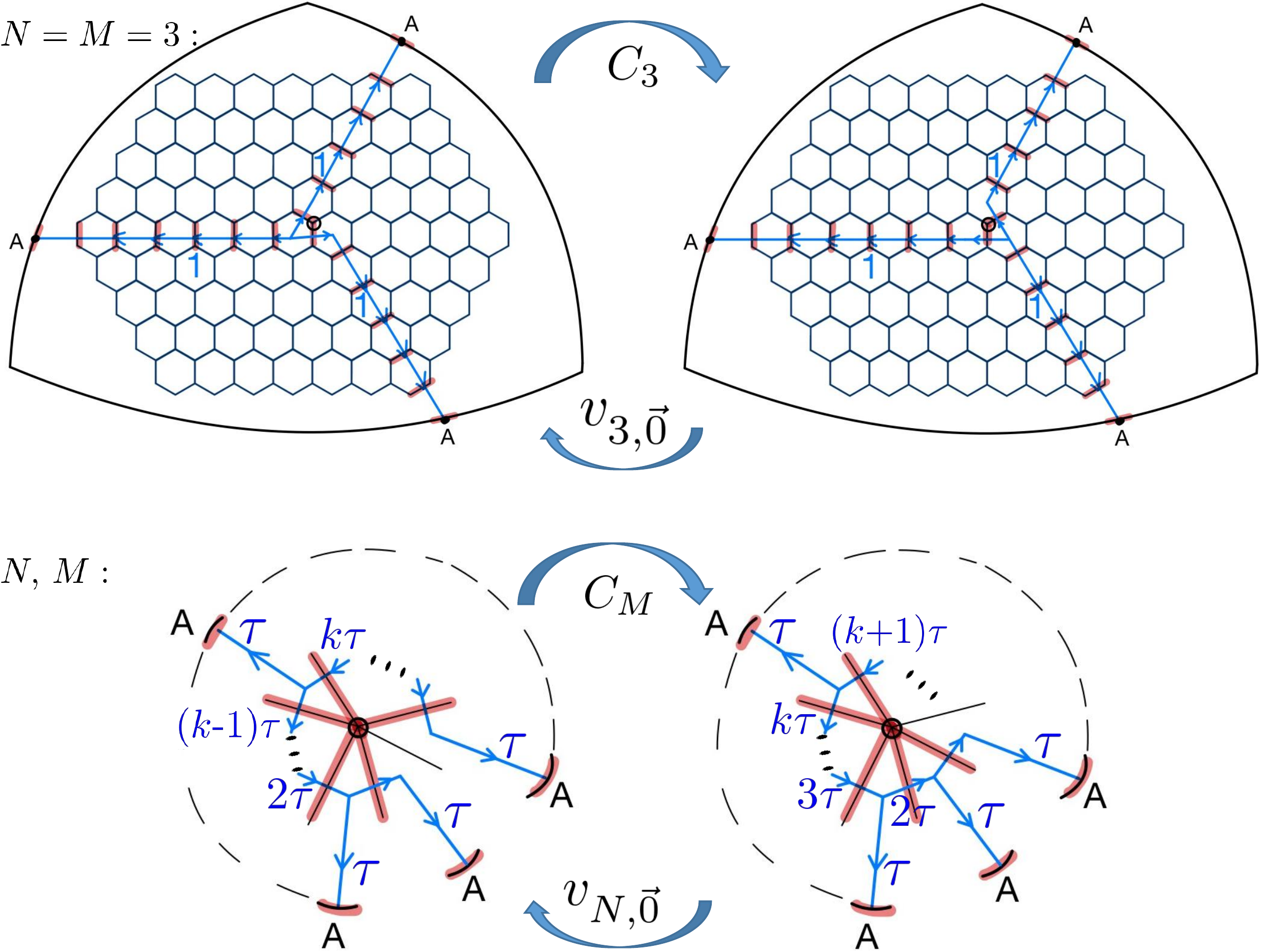}
\caption{The twisted Hamiltonians $\mathcal{H}_\text{tw}$ is presented on the left side, which is transformed to $C_M^{-1}\mathcal{H}_\text{tw}C_M$ on the right side and can be transformed back by $(v_{N,0})^\tau$: $(v_{N,0})^{-\tau}C_M^{-1}\mathcal{H}_\text{tw}C_M(v_{N,0})^\tau=\mathcal{H}_\text{tw}$ with $\tau=N/\text{g.c.d.}(M,N)$.
The twisting numbers should be understood as mod-$N$, so a $p$-twisting is equal to a $(N-p)$-twisting with the opposite arrow direction.  
}
\label{rotation_LSM}
\end{figure}

Let us first consider {an} SU$(3)$ ``spin'' system on a honey-comb lattice respecting $C_3$ and $\mathbb{Z}_3\times\mathbb{Z}_3$ symmetries with a ``spin'' of $b$ at the rotation center $\vec{r}=0$.
{In closing} the boundary,
PBC is incompatible for a similar reason as before, {and}
the only consistent way is to identify or paste the boundary sites by $C_3$ as in FIG.~\ref{PPP_3}~(left).
In the continuum limit,
the resultant space is called a pseudo-projective plane of order $3$ and it has three-fold intersection at the boundary closing,
which is realized at the lattice scale by gluing the three edges together, i.e., boundary-site identifications $\vec{r}\sim C_3(\vec{r})$ as in FIG.~\ref{PPP_3}.
Then we twist the Hamiltonian by $\mathbb{Z}_3${, $V_3$ symmetry,} and deform the twisting configuration as FIG.~\ref{rotation_LSM}~(up-left).
The twisted Hamiltonian $\mathcal{H}_\text{tw}$ preserves $\mathbb{Z}_3\times\mathbb{Z}_3$ symmetry but breaks $C_3$.
Nevertheless,
as shown in FIG.~\ref{rotation_LSM}~(up-right)
it preserves a modified rotation
\begin{eqnarray}
\widetilde{C}_3\equiv C_3v_{N=3,\vec{r}=0},
\end{eqnarray}
which implies a nontrivial commutator between symmetries $\widetilde{C}_3$ and $W_3$:
\begin{eqnarray}
\widetilde{C}_3W_3=\exp\!\left(i2\pi b/3\right)W_3\widetilde{C}_3.
\end{eqnarray}
Thus, when $3$ does not divide $b$, $\mathcal{H}_\text{tw}$ must possess a triply degenerate spectrum and the bulk insensitivity  implies that the original Hamiltonian $\mathcal{H}_0$ cannot have a unique gapped ground state when the central ``spin'' $b$ is not a multiple of $3$.

For general $N$, 
{we consider an SU$(N)$ spin system with $\mathbb{Z}_N\times\mathbb{Z}_N$ and general $C_M$ symmetries, where $M$ and $N$ are unnecessarily equal and there is an SU(N) spin of box $b$ at $\vec{r}=0$.
As shown in FIG.~\ref{rotation_LSM}~(bottom-left), we} close the lattice to form a pseudo-projective plane of order $M$ {and introduce
$M$ lines of $V_N$-twisting to the $\tau$-th power radiating from $\vec{r}=0$ with $\tau\equiv N/\text{g.c.d.}(M,N)$} to have a well-defined twisting, 
i.e., twisting around any plaquette is $0$ mod $N$.
Here ``g.c.d.'' denotes the greatest common divisor.
The modified $\widetilde{C}_M=C_M(V_N)^{\tau}$ is respected by $\mathcal{H}_\text{tw}$ and
\begin{eqnarray}\label{M_N}
\widetilde{C}_MW_N=\exp\!\left(i\frac{2\pi b}{\text{g.c.d.}(M,N)}\right)W_N\widetilde{C}_M.
\end{eqnarray}
Together with the bulk insensitivity,
the above phase factor gives the rotational LSM theorem: when a two-dimensional SU$(N)$ system preserves $\mathbb{Z}_N\times\mathbb{Z}_N$ and $C_M$ symmetries,
a unique gapped ground state is forbidden when the box number $b$ of the ``spin'' at the rotational center is not a multiple of $\text{g.c.d.}(M,N)$.
It explains why $C_3$ and spin-rotation symmetries cannot ensure any ingappability on the spin-$1/2$ honeycomb lattice~\cite{Kimchi-featureless-honeycomb,Metlitski:2018aa} where $b=2s\in\mathbb{Z}$ and $N=2,\,M=3$ coprime.

For example, the SU(4) Kugel-Khomoskii (KK) model with $b=1$ on the square lattice~\cite{Kugel:1982aa,Li:1998aa} is ingappable from Eq.~(\ref{M_N}) with $M=4$ or $2$, which is consistent with the $C_M$-SSB plaquette phase obtained by numerics~\cite{Den-Bossche:2000aa,Mishra:2002aa}.
In contrast, the same SU(4) KK model 
is found to be ingappable (gappable) on the triangular lattice~\footnote{Actually, the triangular lattice has a higher symmetry $C_6$, but it is sufficient to consider its $C_3$ subgroup when discussing the plaquette phase}
from Eq.\ (\ref{M_N}) with $M=2$ ($M=3$), which favors SSB of $C_2$ symmetry while $C_3$ symmetry kept intact; see \cite{Penc:2003aa}.

%\begin{figure}[t]
%\centering
%\includegraphics[width=6.5cm,pagebox=cropbox,clip]{translation_LSM.pdf}
%\caption{The twisted Hamiltonians $\mathcal{H}_\text{tw}$ is presented on the left side, where we suppress the arrows and numbers of twisting for simplicity.
%$\mathcal{H}_\text{tw}$ is transformed by the translation $T_1$ and can be mapped back by a following gauge transformation $v_{N,\vec{C}}$ with $\vec{C}$ indicated above.
%This argument is generalizable to arbitrary dimensions.
%}
%\label{translation_LSM}
%\end{figure}

{\section{Concluding remark}}

Our geometric paradigm suggests {the following} general framework in arbitrary dimensions: 
1) we close the lattice tentatively in all compatible ways with lattice symmetries;
2) we do the symmetry twisting on the bonds transversal to a \textit{non-contractible} {hypersurface of co-dimension} one;
3) the ingappability is extracted {from the algebra of} the modified lattice symmetry.
Thus, it is not restricted to the symmetries {discussed} in this work and we expect that it can be applied to more general settings, e.g., the systems with nonsymmorphic symmetries~\cite{Parameswaran:2013,Watanabe:2015aa} and other rotation-like symmetries such as space dihedral symmetries~\cite{Metlitski:2018aa,Po:2017aa,Else:2020aa}.
{It is applicable to traditional LSM theorems with translations} as shown in Appendix.

\begin{acknowledgments}
The authors thank Masaki Oshikawa, Yasuhiro Tada, and Yunqin Zheng for stimulating discussions.
This work was supported {in part by JSPS KAKENHI (Grant No.\ JP19K03680) and JST CREST (Grant No.~JPMJCR19T2)}.
\end{acknowledgments}

\appendix*
\section{Geometric Derivation of LSM theorems with translation symmetry}
\label{appendix}

\begin{figure}[h]
\centering
\includegraphics[width=6.5cm,pagebox=cropbox,clip]{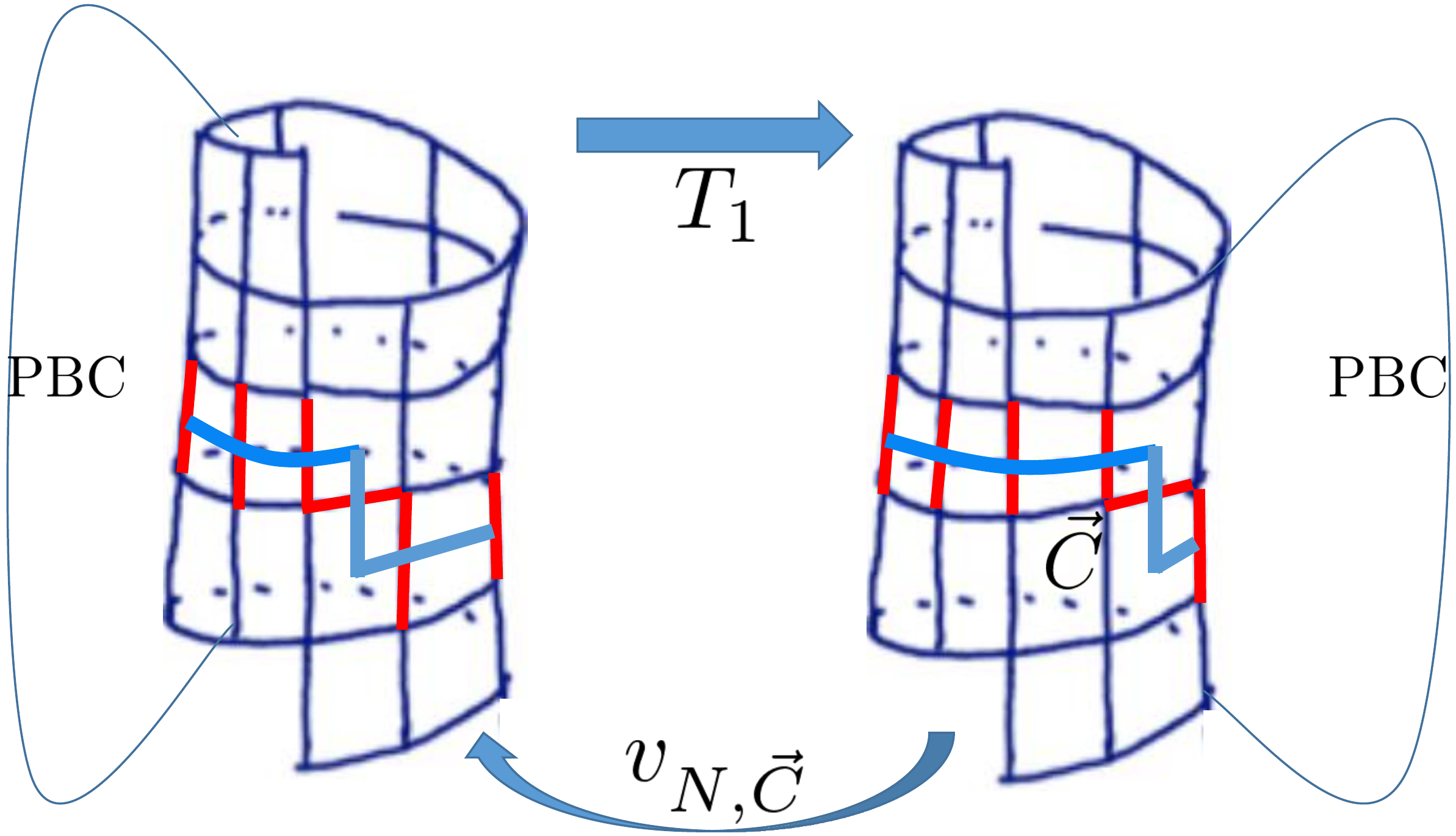}
\caption{The twisted Hamiltonians $\mathcal{H}_\text{tw}$ is presented on the left side, where we suppress the arrows and numbers of twisting for simplicity.
$\mathcal{H}_\text{tw}$ is transformed by the translation $T_1$ and can be mapped back by a following gauge transformation $v_{N,\vec{C}}$ with $\vec{C}$ indicated above.
This argument is generalizable to arbitrary dimensions.
}
\label{translation_LSM}
\end{figure}

In the main text,
we {have focused on LSM-type theorems for lattice Hamiltonians without any translation symmetry imposed.
We have developed a geometric approach using symmetry-twisted boundary conditions to studying} ingappability of spin systems or SU$(N)$ generalizations from rotation, inversion and discrete spin-rotation symmetries.
{In fact,} such a geometric approach can also be applied to traditional LSM theorems with translation symmetries and $b$ boxes per unit cell reviewed in FIG.~\ref{translation_LSM}, where we close the lattice by a tilted boundary condition which is compatible with translations.
The SU$(N)$ symmetry per unit cell at $\vec{r}$ is generated by $v_{N,\vec{r}}$ and $w_{N,\vec{r}}$ satisfying:
\begin{eqnarray}
v_{N,\vec{r}}w_{N,\vec{r}}=\exp\!\left(i\frac{2\pi b}{N}\right)w_{N,\vec{r}}v_{N,\vec{r}},
\end{eqnarray}
where $b$ is $\vec{r}$-independent due to the imposed translation symmetry.
{The} modified translation symmetry $\widetilde{T}_1=T_1v_{N,\vec{C}}$, {where $\vec{C}$ is the spin position imposed by the gauge transformation $v_{N,\vec{C}}$, is preserved by the twisted Hamiltonian and has a commutator with $W_N$ symmetry:}
\begin{eqnarray}
\widetilde{T}_1W_N=\exp\left(i2\pi\frac{b}{N}\right)W_N\widetilde{T}_1.
\end{eqnarray}
{When $b$ is indivisible by $N$, the LSM-type ingappability is concluded from} the bulk insensitivity.

%\sloppy
%\bibliography{bib}


\end{document}